\documentclass{aa}
\usepackage{latexsym}
\usepackage{amsmath}
\usepackage{amssymb}
\usepackage{graphicx}
\usepackage[english]{babel}

\begin{document}

\title{The origin of nitrogen}

\subtitle{Implications of recent measurements of N/O in Galactic metal-poor halo stars}

\author{C. Chiappini\inst{1}, F. Matteucci\inst{2}  \and S. K. Ballero \inst{2}}

\institute{Osservatorio Astronomico di Trieste, via G. B. Tiepolo 11, I - 34131 Trieste, Italia 
\newline e-mail:chiappini@ts.astro.it
\and Dipartimento di Astronomia, Universita' degli Studi di Trieste, via G. B. Tiepolo 11, I - 34131 Trieste, Italia}

\date{Received 1 November 2004 / Accepted 28 February 2005}

\abstract{
Recent new high-precision abundance data for Galactic halo stars 
suggest important primary nitrogen production in very metal-poor massive stars. 
Here, we compute a new model for the chemical evolution of the Milky Way aimed at explaining these new
abundance data. The new data can be explained by adopting: a) the stellar yields obtained from stellar 
models that take into account rotation and b) an extra production of nitrogen in the very metal-poor massive stars. 
In particular, we suggest an increase
of nearly a factor of 200 in $^{14}$N for a star of 60 M$_{\odot}$ and $\simeq$ 40 for a star of 
9M$_{\odot}$, for metallicities below Z$=$10$^{-5}$, with respect to the yields given in the  
literature for Z$=$10$^{-5}$ and rotational velocity of 300 km/s.  
We show that once we adopt the above prescriptions, our model is able to predict 
high N/O abundance ratios at low metallicities and still
explains the nitrogen abundances observed in thin disk stars in the solar vicinity. 
The physical motivation for a larger nitrogen production in massive stars in very metal-poor
environments could be the fact that some stellar models as well as observational data suggest that 
at low metallicities stars rotate faster. 
If this is the case, such large nitrogen production seen in the pristine phases
of the halo formation would not necessarily happen in Damped Lyman-$\alpha$ systems which
have metallicities always above [Fe/H]$\simeq -$2.5, and could have been pre-enriched.
We also compute the abundance gradient of N/O along the Galactic disk and show
that a negative gradient is predicted 
once we adopt stellar yields where rotation is taken
into account. The latter result implies that intermediate mass stars
contribute less to the primary nitrogen than previously thought.

}

   \authorrunning{C. Chiappini et al.}

   \titlerunning{The Origin of Nitrogen}

\maketitle

\section{Introduction}

Investigating the origin of nitrogen in galaxies has been a major
topic of research in the past few years. The reason is at least two-fold: {\it i)} 
There are many processes involved in the computation of the stellar yields
of nitrogen and hence there are still many uncertainties present in these 
calculations (see Meynet \& Maeder 2002a - hereafter MM02). Meynet \& Maeder have
shown that stellar rotation and mass loss can affect the predictions of the stellar yields
especially for He, C, N and O. 
Chemical evolution models can thus be 
used to test and constrain the stellar yields (see Fran\c cois et al. 2004);
{\it ii)} a large amount of data is available for the N/O abundance ratio in different 
environments, ranging from spiral galaxies (e.g. Pilyugin et
al 2004 and references therein) to dwarf galaxies (see Mouhcine \& Contini 2002, Larsen et al. 2001 
and references therein) and damped Lyman alpha systems,
hereafter
DLAs (e.g. Centurion et al. 2003, Prochaska et al. 2002, Pettini et al. 2002). 
In the latter case, the nitrogen evolution 
carries important information on the still-debated nature of these systems.

It has been recently shown (Chiappini, Matteucci \& Meynet 2003, hereafter CMM03) that models
of chemical evolution computed with the MM02 
yields for the whole range of masses, predict a slower increase of nitrogen than what is obtained with
other sets of stellar yields, with important implications for the
interpretation of the DLA abundance data. Due to the slower increase of nitrogen in time,
the DLA abundance patterns can be 
reproduced by ``bursting models'' (see also Lanfranchi \& Matteucci 2003) and in this framework, 
the ``low N/O'' and ``high N/O'' groups of DLAs (first identified by
Prochaska et al. 2002) could be
explained as systems that show differences in their star formation histories
rather than an age difference.
We were able to obtain models that show both a low log(N/O) and a low [O/Fe] 
(of the order of [O/Fe]$\sim$0.2-0.3 dex, in agreement with observations - Centurion et al. 2003) 
during almost all their evolution. Alternative interpretations
(e.g. Prochaska et al. 2002; Centurion et al. 2003)
of the ``low N/O'' DLAs suggested in the literature would imply [O/Fe] ratios larger than
the observed ones. DLAs could also be identified with outer regions of spiral galaxies
(Hou et al. 2001; Calura et al. 2003)
but in this case DLAs with low log(N/O) necessarily would be 
quite young systems (younger than $\sim$150 Myr - see figure 11 of CMM03)
and no discontinuity in the log(N/O) vs. log(O/H) diagram would be expected.

However, as pointed out in CMM03, it remains to be seen to what extent the MM02 yields 
for $^{14}$N in the intermediate mass star range would  
increase once hot bottom burning (HBB) is taken into account.
Although MM02 did not formally include the third dredge-up and HBB, 
it is worth studying the effects of their yields on 
chemical evolution models for the following reasons:
a) The MM02 yields for nitrogen at 
low metallicity result from a new process whose importance 
for chemical evolution has still to be studied.
In the absence of a real quantitative assessment of the 
importance of the HBB it is interesting to 
study the importance of this new process,
which produces ``non-parametric'' yields, independently of HBB
and b) this is particularly justified in view of the fact 
that this new process gives primary nitrogen yields at low 
metallicity not very different from those obtained 
from parametric studies such as van den Hoek 
and Groenewegen (1997 - hereafter vdHG97). This questions the importance
of the HBB\footnote{Marigo (2003)
showed that variable molecular opacities may decrease the efficiency
of HBB - or even prevent it in some cases - especially in the more massive AGB stars}. 
Only by studying the effects separately 
it will be possible to understand the different 
consequences of the two processes.

In the massive range, the yields of MM02 predict some primary nitrogen production
\footnote{We call attention to the fact that
the MM02 yields for helium are currently the only ones to ensure a good agreement between
chemical evolution models for the Milky Way (hereafter MW) and the solar helium abundance - see CMM03. 
This is essentially due to mass loss in massive stars. In the massive range, the yields
computed by MM02 for He, C, N and O would be essentially unchanged
by explosive nucleosynthesis and can thus be considered robust calculations which take into account
important physics (i.e. rotation and mass loss - see Hirschi et al. 2004 for a detailed description of these models for massive stars).}. 
In CMM03 we showed that models for the MW
computed with this new set of yields show
a plateau in log(N/O), due to massive stars with initial rotational velocities of 300 
km sec$^{-1}$, at log(N/O) $\sim-4$. 
This value is below the value of $-$2.2 dex observed in some DLAs and hence we suggested
that in these systems both massive and intermediate mass star, would be
responsible for the nitrogen enrichment 
(in agreement with the conclusions of Chiappini, Romano and Matteucci 2003 and Henry et al. 2000).
This is instead at variance with recent claims that massive stars are the only
ones to enrich systems that show a log(N/O)$\sim-$2.2. However, 
one should keep in mind that stellar evolution calculations for N and O in massive stars
depend strongly on the adopted rotational velocities and mass loss rates, respectively. 

More stringent constraints on nitrogen nucleosynthesis come from the study 
of the nitrogen abundances in stars in the MW since they represent
a true evolutionary sequence, where the stars with lower metallicity are the 
oldest ones (Matteucci 1986). 
Moreover, the halo very metal poor stars play a fundamental
role since, at metallicities below [Fe/H] = $-$3, only Type II supernovae have had time to 
contribute to the interstellar medium enrichment from which these stars formed, thus offering a way
to constrain the nitrogen production in massive stars at low metallicities (the same is true
for other elements as shown by Fran\c cois et al. 2004). 
On the other hand, an important
constraint on the nitrogen production in intermediate mass stars (and thus also on the HBB) 
is the variation of the N/O abundance ratio with galactocentric distance. 
As shown by Diaz \& Tosi (1986), in spiral galaxies, the steepness of the   
abundance gradient of N/O decreases as the primary nitrogen production
in intermediate mass stars increases (see also Chiappini, Romano \& Matteucci 2003).

When we published our two last papers on the evolution of CNO in galaxies
(Chiappini, Romano \& Matteucci 2003 and CMM03)
no conclusive data was available for nitrogen in metal-poor halo stars.
This situation has now greatly changed. 
New data on nitrogen abundances in metal-poor stars (by Spite et al. 2005
and Israelian et al. 2004) show a quite surprising result:
a high N/O ratio suggestive of high levels of production of primary nitrogen 
in massive stars. Moreover, the N/O abundance ratios in metal-poor
stars show a large scatter (roughly 1dex, much larger than
their quoted error bars) although none of the stars measured so 
far has N/O ratios as low as the ones observed in DLAs.

In the present paper we will study the implications of these new data sets for our
understanding on the nitrogen enrichment in our galaxy.
In Sec. 2 we briefly present our model for the MW and the adopted stellar yields. Section
3 is devoted to the comparison between our models for the MW 
and the new data now available for the solar vicinity. 
We will show that currently there is no set of stellar yields able to explain 
the very metal-poor data of Spite et al. (2005). 
Invoking the so-called population III stellar yields available in the 
literature does not solve the problem as it still leads to inconsistencies if one 
considers other abundance ratios, for instance C/Fe. 
This section also includes, for the first time, our predictions for the abundance 
gradients of N/O and C/O once the MM02 stellar yields are adopted. We will show that 
the N/O abundance gradient represents
a powerful tool to assess the importance of HBB in intermediate mass stars. 
In Sec. 4 a discussion is presented where we point out ways to account
for these new observations and check their implications for our previous conclusions
on the nature of DLAs.

\section{Stellar yields and chemical evolution model}

In the present work we adopt the stellar yields described in MM02
and CMM03\footnote{For type Ia SNe we adopted the stellar yields of model W7 of Thielemann et al. (1993).}.  
In Fig. 1 we plot the nitrogen yields of MM02. Filled
squares show stellar yields resulting from models with rotation (V$_{\rm rot}=$300 km/s),
while open symbols stand for models computed with V$_{\rm rot}=$0 km/s. MM02 computed stellar yields
for the following metallicities: Z=0.020 (solid lines), Z=0.004 (short-dashed lines)
and Z=0.00001 (long-dashed lines). The asterisks connected by the long-dashed line show
the stellar yields we adopted for metallicities Z$<$10$^{-5}$ in our heuristic model (see
Section 3).

Two important findings can be seen in Fig. 1: 
{\it i)} in the lowest metallicity case, rotation increases the nitrogen production in stars
of all masses and {\it ii)} the increase of nitrogen, at Z=0.00001, is especially
important in low and intermediate mass stars 
(at least for the case of V$_{\rm rot}=$300 km/s).
However, for other metallicities the $^{14}$N yields of MM02 for the intermediate mass stars
are lower than the ones of vdHG97 as MM02 did not formally include the HBB (see CMM03 their
Fig. 2).

\begin{figure}
\centering
\includegraphics[width=8cm,angle=0]{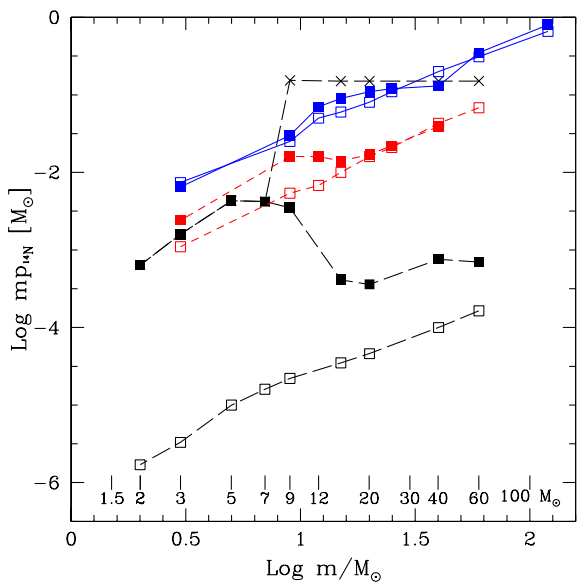}
\caption{MM02 stellar yields for $^{14}$N,
for the whole stellar mass range, for different metallicities. 
The yields of MM02 for stellar
models where rotation is not taken into account are shown as open squares.
Filled squares stand for models with rotation. Stellar yields are shown
for 3 different
values of metallicities (solid lines: solar, dashed line: Z=~0.004 
and long dashed line: Z=~0.00001). The asterisks connected by the long-dashed line
show the stellar yields adopted only for the lowest metallicity case in our
heuristic model (see text)}.
\end{figure}

The adopted chemical evolution model for the MW is the so-called ``two-infall model'' of 
Chiappini et al. (1997, 2001) where a detailed description can be found. 
The fundamental idea of this model is that
the formation of the MW occurred in two different infall episodes,
one forming the halo and part of the thick disk on a relatively short
timescale and another one forming the thin-disk on a longer timescale. 
In this model a threshold gas density
is assumed and, as a consequence, the star formation rate becomes zero
every time the gas density drops below the threshold value. 
The two-infall approach, combined with such a threshold, leads
to a gap in the star formation before the formation of the thin-disk. During 
the ``gap'' in the star formation only elements produced by 
type Ia SNe and low and intermediate mass stars (LIMS), 
born before the ``gap'', are restored into the ISM. As
a consequence this model predicts an increase in the abundance
ratios of elements restored on long-timescales (e.g. Fe or C) 
over $\alpha$-elements (produced mainly by massive short-lived stars) around
a metallicity of [Fe/H]$\sim -$0.6 dex (which corresponds to the time 
of the halt in the SFR which we predict to be around 1 Gyr after the start
of the halo phase - see Chiappini et al. 1997 for details). A star formation halt 
between the formation of the halo and thin disk is suggested by observations
(e.g. Gratton et al. 1996, 2000, 2003; Fuhrmann 1998, 2004).
The required amount of infall
seems to agree well with current estimates and is supported by recent
observations both in our galaxy and in M31 (Sembach et al. 2004 and 
Thilker et al. 2004).

\section{Results}

\subsection{The solar vicinity}

We will concentrate the following discussion on the log(N/O) vs. log(O/H) diagram,
instead of the usual [N/Fe] vs. [Fe/H] one for two main reasons. The first 
one is that since MM02 did not compute stellar yields for Fe it is more consistent
to compare N/O abundance data with our theoretical predictions. The second reason
is that in the case of the sample by Israelian et al. (2004) the uncertainties in N and O
should cancel out once one considers the N/O ratio which is thus less prone 
to observational uncertainties compared to the N/Fe abundance
ratios.

\begin{figure}
\centering
\includegraphics[width=8cm,angle=0]{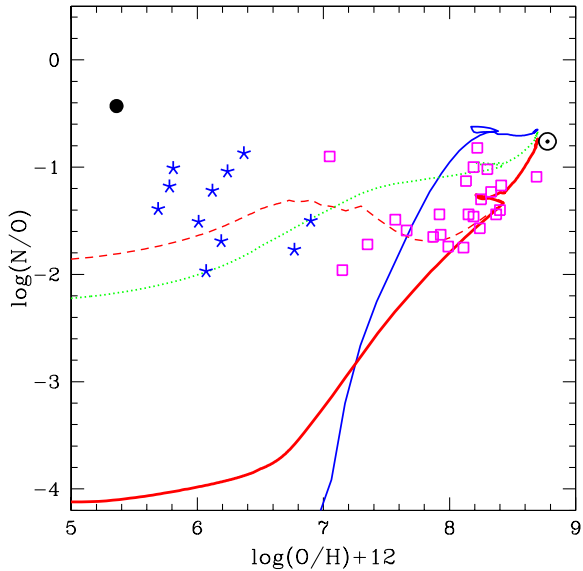}
\caption{Solar vicinity diagram log(N/O) vs. log(O/H)+12. The data points are 
from Israelian et al. 2004 (large squares), Spite et al. 2005 (asterisks). 
Also shown is the very metal-poor star found by Christlieb et al. (2004). Solid
curves show the prediction of MW models computed with vdHG+WW yields (thin solid line) and 
MM02 yields (thick solid line). The latter flattens for log(O/H)+12 $<$ 6.6 due to the contribution
by massive stars to the nitrogen production at low metallicities. The dotted curve shows
a model computed according to the suggestion of Matteucci (1986), where all massive stars,
in all metallicities, contribute a fixed amount of primary nitrogen of 0.065 M$_{\odot}$.
The dashed line shows the prescriptions of our heuristic model computed with MM02 yields but
assuming that massive stars with metallicities less or equal to 0.00001 produce much more nitrogen than
the quantities computed by MM02, as shown
in Fig. 1.}
\end{figure}

One of the main assumptions when comparing chemical evolution predictions with abundance
data is that they represent the pristine abundances from the ISM from which the stars formed.
This means that we should avoid using objects that could have undergone mixing 
processes. Israelian et al. (2004) published homogeneous N/O abundance ratios
for a sample of 31 unevolved dwarf metal-poor stars (shown in Fig. 2 as open squares).

Spite et al. (2005) obtained nitrogen abundances for a sample
of stars of even lower metallicities. Because in this case the abundances were measured in giants,
the authors also measured the surface abundance of lithium as a diagnostic for CNO dredged-up
material to the surface. They were in principle able to select a subsample of ``unmixed'' stars
(plotted as asterisks in Fig. 2). In reality this sample should be seen as an upper limit for N/O
as some mixing could still have taken place. It should also be noticed that although their
abundance sample is unique and reaches a metallicity never reached before (especially for N),
the absolute value of their data points can be still affected by {\it i)} the 3D corrections
applied to their oxygen values and {\it ii)} the fact that the nitrogen abundances were derived 
from the NH band for most of the stars. These abundances show a systematic shift of $+$0.4 dex
with respect to abundances obtained from measurements of the CN band (they had both measurements
for 10 stars - see Spite et al. 2005 for details). Fig. 2 also shows the N/O solar ratio (the solar 
values were taken from Allende Prieto et al. 2001 for oxygen and Holweger 2001 for nitrogen).

Also shown in Fig. 2 is the very
metal poor star (giant) of Christlieb et al. (2004). 
It can be seen that this star has a N/O abundance
ratio which is clearly larger than the typical ratios of the other two data samples.
However, in this case a self-enriched scenario or a contamination by a binary companion 
star still cannot be excluded (see Chrislieb et al. 2004 for a detailed discussion).

In the same figure we show our model predictions for different assumptions of stellar yields.
The solid curves are the same models shown in CMM03 (their figure 7): the thin line
represents a model computed with the stellar yields of vdHG97 and 
Woosley \& Weaver (1995 - hereafter WW95),
the thick line shows a model computed with MM02 stellar yields.
When comparing these two models two things can be noticed
: a) as WW95 do not produce primary
nitrogen in massive stars the thin curve computed with their stellar prescriptions 
does not flatten at low metallicities (contrary
to what happens to the thick curve because in this case, according to the prescriptions
of MM02, the massive stars produce some primary nitrogen)
and b) the increase of the N/O ratio as a function of metallicity
in the model represented by the thin curve is faster than the one shown by the thick curve. 
This is mainly due to the large amount of $^{14}$N produced during the HBB in intermediate mass stars 
according the calculations of vdHG97. 

Before discussing the other two models plotted in Fig. 2, 
notice that the new points of Israelian et al. (2004) are not far from the thick solid curve
computed with the MM02 stellar yields, especially for log(O/H)+12 $>$ 8.0. 
As discussed in the previous section, because MM02 did not formally include HBB, one 
would expect the curve to lie much below the data points. The fact that the thick curve is 
close to the abundance ratios measured by Israelian et al. (2004) in unevolved stars suggests
that HBB is less efficient than in vdHG97 models\footnote{Here we adopted their standard
models and took their tables where the mass loss parameter varies with metallicity - see CRM03 for details. 
vdHG97 also computed another set of models where less HBB was assumed. 
We also computed a chemical evolution
model where the latter stellar prescriptions were adopted. In Fig. 2 this model would fall in between the two
solid curves discussed here - it is not shown to make the figure less crowded.}. 
In fact, the thin curve lies 
above most of the Israelian et al. (2004) data points\footnote{
Romano \& Matteucci (2003) have shown that by adopting the vdHG97 set of 
stellar yields with less HBB it is possible to reproduce the trend of $^{12}$C/$^{13}$C, 
which decreases in time in the solar neighbourhood.}.

The dotted curve was computed according a suggestion made by Matteucci (1986) that 
all massive stars should produce around 0.065 M$_{\odot}$ of primary
nitrogen (it is the same model shown by the thick solid curve in Fig. 2 except that in this 
case we assume that all massive stars, for all metallicities, contribute a fixed
amount of 0.065 M$_{\odot}$ of N).
This suggestion was based on the little data available at that time which seemed
to suggest a flat [N/Fe] ratio at low metallicities. As it can be seen, the dotted
line can reproduce the locus of the Spite et al. (2005) data sample but tends to overproduce
nitrogen at higher metallicities (this curve is above most of the Israelian et al. data,
even though it can reproduce the solar N/O abundance ratio).

The dashed line shows what we call our heuristic model. This model is the same as the 
thick solid line except that for the metallicities Z $<$ 0.00001 we increased the 
yields of nitrogen given by MM02 for massive stars in the following way:
we added 0.15M$_{\odot}$ of nitrogen in the table of MM02 for Z=0.00001
(which translates into a factor of 200 increase in $^{14}$N for a 60 M$_{\odot}$ star and around a factor
of 40 for a 9M$_{\odot}$ star). This is shown in Fig. 1 by the asterisks connected by a long-dashed line.
As our code then interpolates (linearly) the stellar yields for metallicities
between Z=0.0 and Z=0.00001, this model produces more nitrogen
at the beginning of galaxy evolution, leading to large N/O ratios at low metallicities
but not changing its behavior for metallicities more close to solar (in fact the dashed curve
coincides with the thick curve for oxygen abundances above 8.0).

The physical motivation for this heuristic model would be an increase
of the rotational velocity in very metal-poor stars. 
As shown by MM02, the nitrogen yields increase with increasing rotational velocities.
It might be that the initial
distribution of the rotational velocities is different at different
metallicities. There are some indirect indications that, at low 
metallicities, there are more fast rotators. For instance, the observed
fraction of Be stars (which, being very fast  rotators, are near the break up limit) 
appear to be more frequent in the SMC than in the MW (Maeder et al. 1999). 
Keller (2004) found that the rotation velocities of early B-type stars in the
LMC are higher than the rotation velocities of comparable stars in the MW.
Part of the reason for this is that a zero metallicity 
star having the same amount of angular momentum as
a solar metallicity star rotates much faster due to its greater
compactness (see Meynet \& Maeder 2002b).
Thus it might be that only stars at low metallicity rotate sufficiently
fast to enable massive stars to contribute large amounts of nitrogen.
Whether these suggestions are physically
plausible remains to be assessed by future stellar evolution models, including rotation and 
mass loss.

If the nitrogen production in very-metal poor massive stars depends strongly
on the rotational velocity of the star, this could explain
the large scatter observed in N/O at low metallicities. Moreover,
the scatter could be related to the distribution of the stellar rotational
velocities as a function of metallicity (being more biased to larger values
as the metallicity decreases). Clearly,  the above suggestion needs to be confirmed 
but it can in principle be tested by observations.

\subsection{Is there an alternative explanation to rotation ?}

An alternative way to explain the high N/Fe ratios at low metallicities
is to assume that the first stars to enrich the ISM were population III stars (hereafter PopIII). 
Recently, Akerman et al. (2004) found an upturn in C/O at low O/H, 
and suggested that this could also be explained 
by adopting  PopIII stellar yields (see figure 8 of Akerman et al. 2004). 
In particular, they adopted 
stellar yields computed by Chieffi \& Limongi (2002) for metal-free
supernovae which, according to the latter authors, should be C-rich\footnote{Notice that Chieffi \& Limongi (2002) do not include rotation in their computations.} and assumed that PopIII stars were born with 
a ``top heavy'' IMF. Soon afterwards, Spite et al. (2005) confirmed that the
upturn in C/O at low O/H values found by Akerman et al. (2004) extends
to lower metallicities (see Fig. 3, where the open squares show the abundances measured
by Akerman et al. 2004 and the asterisks show the Spite et al. 2005 data). 

In this section we will check if current stellar yield calculations 
for the so-called PopIII stars are able to ensure a good 
fit of the log(N/O) vs. log(O/H) diagram
and, at the same time, still explain the almost flat behaviour
of [C/Fe] found by Spite et al. (2005). The only set of stellar yields
for PopIII able to produce a high enough N/Fe 
at low metallicities is the one of Chieffi and Limongi (2002,2004) (for a more detailed
discusssion on the role of PopIII stars in the ISM enrichment of several
other elements and models adopting different
prescriptions for PopIII stars, see Ballero et al. 2005).

Here we find that although the latter stellar yields can explain the C/O vs. O/H 
upturn at low metallicities 
they fail to reproduce the almost flat [C/Fe] abundance ratios found to extend
down to the low metallicities sampled by Spite et al. (2005).
This can be clearly seen in Figures 3 and 4 where the dot-dashed lines show a model similar 
to the one of Akerman et al. (2004), i.e. a model where the
prescriptions for PopIII stellar yields of Chieffi \& Limongi (2002, 2004) are adopted
for $Z <$10$^{-6}$. The latter value corresponds to the threshold 
metallicity ($\simeq$ 10$^{-4}$Z$_{\odot}$) below which the IMF should be ``top-heavy'', as
suggested in the literature for the so-called PopIII stars (where PopIII stands not only
for ``zero metallicity stars'' but also stars born with a different IMF, where low and intermediate
mass stars did not form - see Ballero et al. 2005 for details). Therefore,
this model was computed following the Akerman et al. (2004) prescriptions also for the IMF, i.e.
a ``truncated'' Scalo (1986) IMF, with M$_{low}$ $=$ 10 M$_{\odot}$ for $Z <$10$^{-6}$ 
and a normal Scalo (1986) IMF for higher metallicities. The dot-dashed model can
well explain the C/O vs. O/H observations but leads to C/Fe ratios that are above the 
observed values (see dot-dashed curve in Fig. 4, 
upper panel - the data are from Cayrel et al. 2004) and is not able to produce the 
amount of N required at low metallicities to explain
the new data points of Spite et al. (2005 - see dot-dashed curve in Fig. 4, bottom panel).

\begin{figure}
\centering
\includegraphics[width=8cm,angle=0]{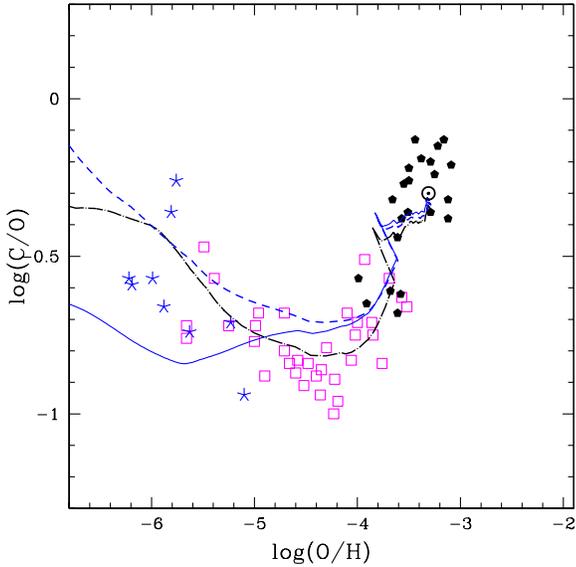}
\caption{Log(C/O) vs. log(O/H) diagram. The data are from Spite et al. (2005 - asterisks),
Israelian et al. (2004 - squares), Nissen (2004 - filled pentagons). The different curves
show our model predictions computed with different stellar yields as follows: 
a) dot-dashed curve - a model computed according the prescriptions of Akerman et al. (2004);
b) thin solid curve - vdHG97 and WW95, where the latter refer to their solar tables and c) dashed line - vdHG97 and WW95 (where in this case the oxygen as a function of metallicity was adopted as suggested
by Fran\c cois et al. 2004). This figure shows an alternative to the model suggested by Akerman et al. (2004) to obtain an upturn of C/O at low
metalliticies.}
\end{figure}

\begin{figure}
\centering
\includegraphics[width=8cm,angle=0]{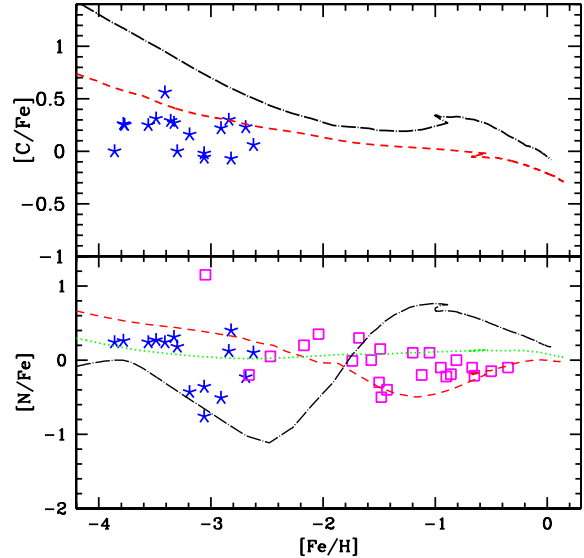}
\caption{The data points are from Cayrel et al. (2004), Spite et al. (2005) (asterisk)
and Israelian et al. (2004) (squares). 
The dashed line shows the predictions of our ``heuristic model''.
The dot-dashed lines refer to models computed with Chieffi \& Limongi (2002,2004) for metallicities
below 10$^{-6}$ and a top-heavy IMF (see text). 
In the lower panel a model that assumes a constant N production in massive
stars of all metallicities (Matteucci 1986) is also shown. In this figure the models are normalized
to the solar N and Fe of Holweger (2001) and to the C of Allende-Prieto et al. (2001)}
\end{figure}

Also shown in Fig.4 is our heuristic model (dashed-line). This model can well explain
the C/Fe ratios found in very metal-poor stars by Cayrel et al. (2004) (as expected since
our heuristic model is identical to the CMM03 model as far as C and Fe are concerned)
and also provides a good agreement 
with the N/Fe abundance ratios of Spite et al. (2005). However, in this case our heuristic
model predicts a "valley" at intermediate metallicities (although less
pronounced than the one shown by the dot-dashed model where the PopIII contribution
was taken into account), whereas the 
data show a more flat behavior of N/Fe for the whole metallicity
range. An easy way to obtain a flatter curve is to assume
that the low and intermediate mass stars at low metallicities also should have higher
nitrogen yields than the ones computed for Z$=$10$^{-5}$. It is not excluded 
that intermediate mass stars of that same low metallicity could also produce large amounts
of nitrogen if this production is linked to high rotational velocities (as discussed
in Sect. 3.1). However, in this paper we wanted to change as little as
possible the already-existing stellar yields of MM02 and thus we increased
the nitrogen yields only for massive stars. Our main goal here is to explain the very
metal-poor data of Spite et al. (2005) and for this we need massive stars because intermediate
mass stars would not have had time to contribute to the ISM enrichment at 
such low metallicities, given their longer lifetimes. In Fig. 4 we also show a model in which
we assumed a constant N production in massive stars of all metallicities (Matteucci 1986 - dotted
curve, also shown in Fig. 2. In this case a flat N/Fe vs. Fe/H is also obtained (see Ballero et al. 2005).

It is beyond the scope of the present paper to discuss in detail the problems related to
carbon nucleosynthesis (for that see our previous papers CMM03 and Chiappini, Romano \& Matteucci 2003), 
but Fig. 3 illustrates that, for the low metallicity end, an upturn in C/O can also be obtained without
the need to invoke PopIII stellar yields/IMF. In this figure
the solid thin curve shows model 7 of Chiappini, Matteuci
and Romano (2003). The dashed curve shows the same model but adopting the WW95 
stellar yields of oxygen as a function of metallicity 
(as suggested by Fran\c cois et al. 2004 and Goswami \& Prantzos 2000).
In this case a C/O upturn can be obtained at low metallicities. This is because
WW95 predict a decrease in oxygen rather than an increase in carbon for Z=0. 
As a consequence, such a model still fits the [C/Fe] abundance
ratios as a function of metallicity 
\footnote{Some mechanism able to increase
$^{14}$N at low metallicities is still needed in this case as models computed with WW95 stellar yields are not
able to fit the new Spite et al. (2005) data for N/O, as shown by the thin solid curve in Fig. 2. 
WW95 also did not include
rotation in their calculations.}). Notice that in this 
case the model computed with the stellar yields of MM02 
cannot fit the C/O observations either, although they provide a good 
fit for [C/Fe] (see CMM03, their Fig. 6).
The above results illustrate the importance of testing PopIII stellar yield
predictions simultaneously on different abundance ratios (see Ballero et al. 2005
for a discussion of several other abundance ratios). 

In summary, our results suggest 
that a large $^{14}$N yield in massive stars is required to fit the new abundance
data for very low metallicities. Rotation
seems to be the most promising way to explain the new data and its scatter, whereas current
PopIII stellar yields able to produce a high N/Fe at low metallicities tend to overproduce
carbon at variance with the flat C/Fe vs. Fe/H observed . Current stellar models
that take into account rotation (MM02) do not provide the required amount of nitrogen
to fit the data. However, it is worth noticing that MM02 computed stellar 
yields down to Z$=10^{-5}$. It has to be seen if computations to even lower
metallicities will be able to produce more N, at the levels suggested by our results.
Another alternative could be an enhanced nitrogen production in massive close 
binaries (see Wellstein et al. 2001, Langer 2003 and references therein).

However, if future stellar evolution models for very low initial metallicities are 
able to produce large amounts of nitrogen, it should still be checked to what extent C would also
be produced. A large production of C at low metallicities would make it difficult 
to explain the flat behaviour observed in [C/Fe] from solar metallicities to [Fe/H] as low as $-$5.
The results discussed above suggest that a better agreement in all plots would be obtained
if the low-Z calculations were able to simultaneously increase the stellar 
yields of $^{14}$N, keeping C almost unchanged and decreasing the stellar yields of oxygen.

\subsection{Present abundance gradients}

In this last section we check the effect of the MM02 stellar yields on
the abundance gradients predicted for our galaxy. As shown by Prantzos (2003),
not much difference is seen when plotting the C/H, N/H and O/H abundances as 
a function of the galactocentric distance for models computed with WW95+vdHG97 or
MM02 stellar yields. We confirm this result. However, as shown by Chiappini, Romano \& Matteucci
(2003) this is not the case for the N/O, C/O and C/N abundance ratios. Figure 5 shows our 
predictions for the variation of these abundance ratios as a function of galactocentric
distance obtained with MM02 stellar yields (thick lines) compared with model 7 of 
Chiappini, Romano \& Matteucci (2003 - thin line), which
was computed with WW95+vdHG97 stellar yields. Important differences can be seen when the different
sets of yield are adopted.

\begin{figure}
\centering
\includegraphics[width=8cm,angle=0]{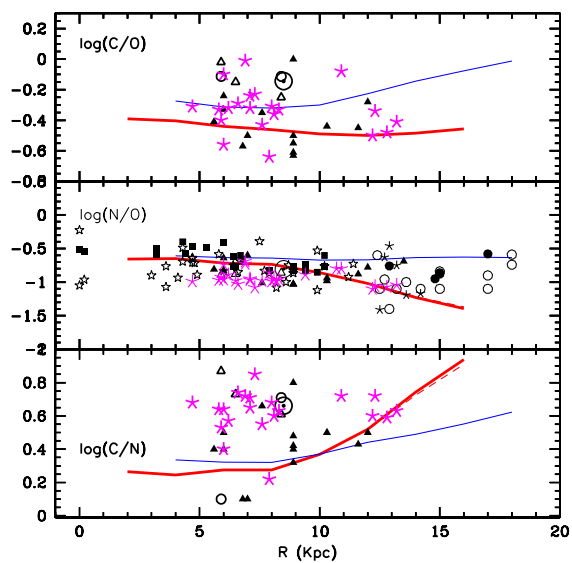}
\caption{Abundance gradients of C/O, N/O and C/N predicted by
models adopting vdHG+WW yields (as model 7 of Chiappini, Romano \& Matteucci 2003 - thin solid line)
and the same models computed with MM02 yields (as described in CMM03 - thick solid line). 
The dashed line barely seen in the bottom diagram corresponds
to the predictions of our heuristic model (see text). 
In the middle panel both models overlap. Therefore, it is clear 
that the dominant
factor of the N/O gradient in the MW is nitrogen production in LIMS and not the primary
nitrogen in massive stars. For the abundance data see Chiappini, Romano \& Matteucci (2003) and 
references therein. Here we added the recent abundance data of Daflon \& Cunha (2004 -
large asterisks).}
\end{figure}

In Fig. 5 we see a large scatter in the data points, especially
for C/O abundance ratios (upper panel). Moreover, in the upper panel, where the open symbols stand for HII regions
and the filled symbols and asterisks represent B stars, it can be seen that the latter tend to show
systematically lower C/O abundance ratios. More data is necessary to use these
abundance gradients as tools to better constrain the carbon and nitrogen nucleosynthesis 
(especially in intermediate-mass stars). 
We recall (see Chiappini, Romano \& Matteucci 2003) that the C/O predictions shown
by the thin curve were computed with vdHG97 yields for the case where the mass loss parameter increases
with metallicity.
This leads to a larger C production at lower metallicities\footnote{Lower mass loss rates 
lead to longer stellar lifetimes. As a consequence
the star undergoes more dredge up episodes thus increasing the amount of C brought to the surface and later
ejected into the interstellar medium.}. As a consequence this model predicts an increase of the C/O abundance
ratio towards the outer parts of the disk, where the
contribution of low metallicity stars dominates. 
The model computed with MM02 stellar yields leads
to a flat C/O abundance ratio along the disk (see thick solid curve in the upper panel of Fig. 5). 
This is because, due to the lack of the third dredge
up, intermediate-mass stars contribute in a negligible way 
to the abundance gradients and what is seen in this case
is essentially the result of the enrichment due to massive stars. This also 
explains why the absolute C/O abundance ratios in this case are systematically lower than in the 
thin-curve model (see Henry 2004 and CMM03).

The middle panel of Fig. 5 shows a very interesting result: a model computed with vdHG97+WW95
stellar yields (thin curve) leads to a flat N/O gradient, whereas the model computed with MM02
(thick curve) leads to a decrease of the N/O abundance ratio as a function of the galactocentric distance.
This is a very important result. As discussed in Chiappini, Romano and Matteucci (2003), 
other galaxies, like M101, clearly show a negative N/O gradient. 
In that paper we showed that if the vdHG97
stellar yields were adopted it was impossible to obtain a negative gradient for M101 and attributed
this to the fact that vdHG97 predict too much nitrogen in intermediate-mass stars (due to a very efficient HBB). 
The fact that the curve computed with the MM02 stellar yields leads to a negative gradient for N/O in the MW 
(still consistent
with the data\footnote{Optical data suggest a flat N/O 
abundance gradient for the MW, whereas infrared data suggest a negative one (Simpson et al. 1995). Moreover, the recent data by Daflon \& Cunha (2004) also show a gradient in N/O very similar 
to the one obtained by our model once the MM02 stellar yields are adopted (compare asterisks and thick line shown in the middle panel of Fig. 5).}) again suggests that the quantity of nitrogen ``missing'' in their calculations, due to the
fact that these authors do not include the HBB, should be small. 

Negative abundance gradients for N/O have been observed in many other spiral galaxies as shown
recently by a compilation of more than 1000 published spectra of HII regions in spiral galaxies by
Pilyugin et al. (2004). Previous determinations
of O/H abundance gradients (e.g. Diaz et al. 1991) in galaxies could 
have been overestimated in inner disks (see Garnett et al. 2004), 
which would lead to flatter N/O abundance gradients.

In the middle and lower panels of Fig. 5 we also plotted our heuristic model (which is essentially
like the thick curve but where we increased the yields of $^{14}$N in massive stars at low metallicities - 
see previous Section). This model (dashed curve) can be barely seen as it almost overlaps 
with the thick line model. This shows
that the abundance gradients in the MW depend on the stellar yields in intermediate mass stars 
(as the 
metallicities do not reach the low values seen in Fig. 2 or 3 even in the outermost parts of 
the galactic disk - see also Diaz \& Tosi 1986).

\section{Discussion and conclusions}

In this paper we computed chemical evolution models for the MW aimed
at explaining the new nitrogen abundances measured recently in halo stars 
(Spite et al. 2005 and Israelian et al. 2004). In particular,
we computed what we call our heuristic model for the MW
where nitrogen stellar yields of massive stars were increased only for the lowest metallicity
with respect to the ones published by MM02. 
Our main conclusions are:
\begin{itemize}

\item A mechanism able to 
produce more $^{14}$N in massive stars at low metallicities relative
to the existing stellar yields is necessary in order to explain
the new data.
If this large nitrogen production is linked to the fact that, at low metallicities, stars should in principle
rotate faster (as discussed by Meynet  \& Maeder 2002b) it would also offer a way to explain
the scatter in N/O measured at these metallicities. 

\item To also reproduce the observed
abundances of C/O and C/Fe in Galactic halo stars, it is important that the production of 
primary nitrogen in massive stars at metallicities below Z$=$10$^{-5}$ is accompanied by a decrease in
oxygen and almost no change in carbon. Whether these suggestions are physically
plausible is still to be assessed by future stellar evolution models, including rotation and 
mass loss.

\item 
Rotation in intermediate mass stars is also able to produce primary nitrogen. We show that even if MM02 did not formally include the HBB, models computed with their stellar yields are not far from the 
abundance data in the solar vicinity 
(in the metallicity range where the IMS are supposed to contribute) and
are still compatible with the abundance gradient for N/O along the Galactic disk. 
Although the data for the MW 
are not yet conclusive about the existence of a N/O abundance gradient, abundance gradients
are clearly observed in other spiral galaxies. The existence of abundance gradients of N/O in spiral galaxies imposes limits on the efficiency of HBB since for high efficiencies the gradients would vanish 
(see also Chiappini, Romano \& Matteucci 2003).

\end{itemize}

If the new case presented here (shown by the dashed curve in Fig. 2)
is accurate then it might be that only stars at such low metallicities rotate sufficiently
fast to enable massive stars to contribute large amounts of nitrogen. 
If this is the case, our interpretation of the two DLA groups observed in the N/O vs. O/H 
diagram as 
being the result of different star formation histories rather than an age
difference (given in CMM03) would still be possible: it could be that
in DLAs the ISM was never as metal poor as the one from which the halo stars studied by 
Spite et al. (2005) formed. In fact, 
DLAs show metallicities higher than [Fe/H] $\simeq -$2.5. This could happen
if, for instance, the ISM in DLAs suffered a pre-enrichment phase before
the start of star formation. 

This is easier to envisage in the case of 
outer spiral disks as progenitors of DLAs. 
As shown by Chiappini et al. (2001), the outer parts
of spiral disks could have been pre-enriched by halo/thick disk gas.
If this is the case, the large
nitrogen production seen in halo stars would not necessarily have 
taken place in DLAs. In other words, 
in DLAs very fast rotating massive stars probably never existed and this explains why
these systems still show the lowest N/O ever measured. 

Although the data analyzed here are the best at 
currently available, there is still the possibility that the so-called ``unmixed
stars'' receive a minor contribution from CNO processing material and that
the nitrogen abundance could have been overestimated. 

\begin{acknowledgements}

We would like to thank F. Calura, S. Recchi, D. Romano and G. Meynet for their
suggestions on an earlier draft. C.C. and F.M. acknowledge
financial support from the Italian MIUR (Ministery for University and Scientific
Research) through COFIN 2003, prot. 2003028039. We also thank the 
referee, Dr. Argast, for his insightful comments that helped to improve this work.
 
\end{acknowledgements}

\end{document}